\documentclass[fleqn,usenatbib,useAMS]{mnras}
\usepackage[utf8]{inputenc}
\usepackage{ae,aecompl}

\usepackage{graphicx}	
\usepackage{epstopdf}
\usepackage{amsmath}	
\usepackage{amssymb}	

\def\apj{ApJ}
\def\apjl{ApJL}
\def\aap{A\& A}

\def\mnras{MNRAS}

\def\apjs{ApJS}

\def\mnras{{MNRAS}}
\def\pasj{{PASJ}}

\def\ala#1{$^{#1}$}
\def\alamenos#1{$^{-#1}$}

\newcommand{\alphavir}{\alpha_{\rm vir}}

\newcommand{\Msun}{$M_\odot$}    %
\newcommand{\Msunmath}{M_\odot}    %

\newcommand{\Npdf}{$N$-PDF}    %

\newcommand{\Rtrans}{R_{\rm trans}}
\newcommand{\Sigmath}{\Sigma_{\rm th}}

\title[Larson Mass-size relation]{What is the physics behind the Larson mass-size relation?}

\author[J. Ballesteros-Paredes et al.]{J. Ballesteros-Paredes$^{1}$\thanks{E-mail: j.ballesteros@irya.unam.mx (JBP)},  C.  Rom\'an-Z\'u\~niga$^{2}$, Q. Salomé$^{1}$,   M. Zamora-Avil\'es$^{3}$ and 
\newauthor
M. J. Jiménez-Donaire$^4$
\\
\\
$^{1}$Instituto de Radioastronom\'ia y Astrof\'isica, UNAM, campus Morelia. PO Box 3-72. 58090. Morelia, Michoac\'an, M\'exico
\\
$^{2}$Instituto de Astronom\'ia, UNAM, campus Ensenada. Ensenada, Baja California. 22860. México
\\
$^{3}$CONACYT-Instituto Nacional de Astrof{\'i}sica, {\'O}ptica y Electr{\'o}nica, Luis E. Erro 1, 72840 Tonantzintla, Puebla, M{\'e}xico
\\
$^4$Harvard-Smithsonian  Center  for  Astrophysics,  60  GardenStreet, Cambridge, MA 02138, USA; 
}

\date{Accepted XXX. Received YYY; in original form ZZZ}

\pubyear{2019}

\begin{document}
\label{firstpage}
\pagerange{\pageref{firstpage}--\pageref{lastpage}}
\maketitle

\begin{abstract}

{Different studies have reported a power-law mass-size relation $M \propto R^q$ for ensembles of molecular clouds. In the case of nearby clouds, the index of the power-law $q$ is close to 2. However, for clouds spread all over the Galaxy, indexes larger than 2 are reported. We show that indexes larger than 2 could be the result of line-of-sight superposition of emission that does not belong to the cloud itself. We found that a random factor of gas contamination, between 0.001\%\ and 10\%\ of the line-of-sight, allows to reproduce the mass-size relation with $q \sim 2.2-2.3$ observed in Galactic CO surveys. Furthermore, for dense cores within a single cloud, or molecular clouds within a single galaxy, we argue that, even in these cases, there is observational and theoretical evidence that some degree of superposition may be occurring. However, additional effects may be present in each case, and are briefly discussed. We also argue that defining the fractal dimension of clouds via the mass-size relation is not adequate, since the mass is not {necessarily} a proxy to the area, and the size reported in $M-R$ relations is typically obtained from the square root of the area, rather than from an estimation of the size independent from the area. Finally, we argue that the statistical analysis of finding clouds satisfying the Larson's relations does not mean that each individual cloud is  in virial equilibrium.
}
\end{abstract}




\section{Introduction}\label{sec:intro}

Nearly 40 years ago, \citet{Larson81} found, for an ensemble of clouds, three empirical relationships between their basic properties: (a) a power-law scaling between size $R$ and the velocity dispersion $\sigma_v$, $R\propto \sigma_v^{1/2}$; (b) an approximate equipartition between their kinetic and gravitational energies, implying that clouds are gravitationally bound, and (c) a power-law scaling between the density $n$ and size, $n\propto R^{-p}$. The first relation has been traditionally interpreted as evidence of compressible turbulence \cite[e.g., ][and references therein]{BP+07}. However, as it has been possible to get measurements of such properties in regions with larger column density environments, it has been found that there is not a single velocity dispersion-size relationship with a clear exponent, but a scatter plot \citep{BP+11a, BP+18}. This fact has been interpreted by these authors as evidence of gravity driving the kinetic motions in molecular clouds (MCs). These results, furthermore, agree with the second relation: if gravity drives the kinetic motions, it is natural to expect MCs to follow the second relation, i.e., to be in  nearly energy equipartition\footnote{Energy equipartition has also been interpreted as virial equilibrium. However, there are serious differences between one concept and the other as discussed by \citet{BP06}.}  \citep{HBB01, VS+07}.

The third relation implies that the column density $N = n\ R$ of a set of clouds with a wide variety of masses, sizes and evolutionary states, is nearly constant. It also implies a mass-size correlation of the form

\begin{equation}
    M \sim n\ R^3 \propto R^q,
    \label{eq:MR}
\end{equation}
with $q = 3-p$.  At face value, Larson's data imply $q = 1.9$. Using the best estimations of mass available provided by extinction maps, \citet{Lombardi+10} found one of the more tight correlations in astronomy ever reported: $M\propto R^2$, but studies using CO data systematically estimate larger slopes. For instance, 
\citet[e.g., ][]{Roman-Duval+10} found values of $q\sim 2.2$ for first quadrant clouds, while \citet[][]{Miville-Deschenes+17}, found $q\sim 2.36$, using the data from the \citet[][]{Dame+01} survey of the whole Galaxy. 
All these studies also show substantially more scatter than the extinction studies. However, the differences between the results using dust extinction and CO data are significant.  In the present work we focus on the origin of the multi-cloud mass-size relation, in an attempt to explain such differences.  We will discuss the origin of the correlation, and the discrepancy between those using  extinction data for nearby clouds, and the results using CO data for clouds spread all around the Galaxy. {In \S\ref{sec:explanation} we discuss why a slope of $q=2$ in the mass-size relation should be expected for clouds that do not overlap, and why superposition effects should increase this slope. In \S\ref{sec:model} we propose a simple model in which some arbitrary amount of gas in the line of sight of the observed cloud can change the slope of the mass-size relation, and show that indeed, a random fraction between $10^{-4}$ and 0.1 of the mass in the line of sight may produce a mass-size relation with slope of $q\sim 2.3$. In \S\ref{sec:discussion}, we discuss our results, and argue what happens either in the case of cores in single clouds, and clouds in single galaxies
We also call into question whether the mass-size relationship should be used for computing the fractal dimension of cores. Finally, in \S\ref{sec:conclusion} we provide our conclusions.}

\section{The underlying physics in the mass-size relation}\label{sec:explanation}

\begin{figure*}
     \includegraphics[width=2\columnwidth]{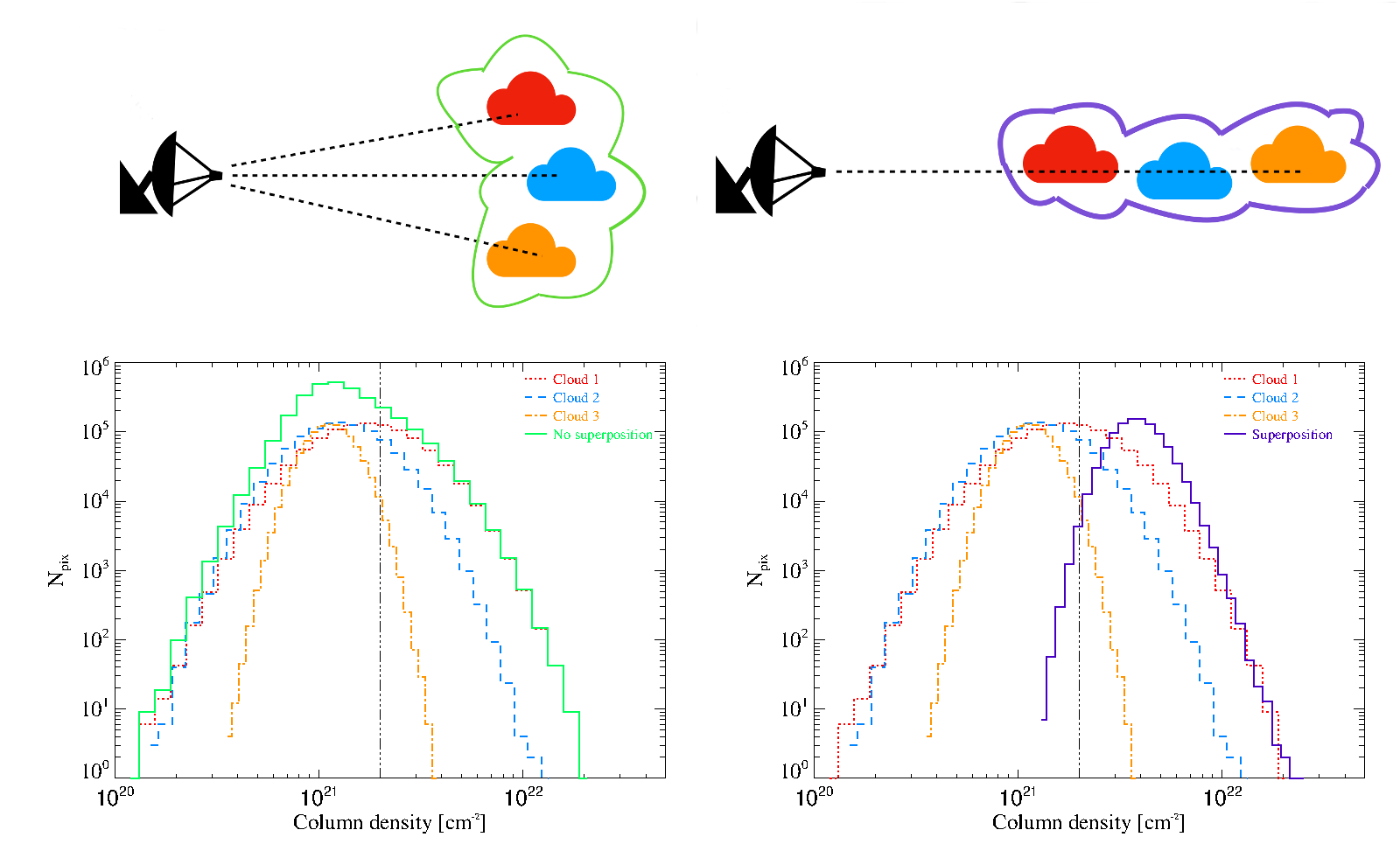}
    \caption{ {Upper left panel: a schematic view of three { unaligned} molecular clouds, which are seen as one molecular complex due to resolution limitations. Lower left panel: the resulting \Npdf\  (green histogram), which is just the addition of the individual \Npdf s (red, blue, and orange lines). The vertical dash-double dotted line represents a threshold assumed to define clouds. Right panels: Same as left, except that the three individual clouds are now aligned along the line of sight. The purple histogram in the lower right panel is the \Npdf\ from the superposition of the individual \Npdf s.}}
    \label{fig:Npdf_dibujitos}
\end{figure*}

By defining clouds as connected sets of pixels above a particular column density threshold\footnote{Molecular gas is dissociated at column densities below $N\sim 10^{21}$cm\alamenos 2 \citep[see, e.g., ][]{HBB01}.}, the average column density for a molecular clouds necessarily close to the value of the threshold, just because the vast majority of points in each cloud is close to that value. This point was noticed first by \citet[][]{BM02}, using numerical simulations of molecular clouds, and discussed in more detail from an analytic perspective by \citet[][]{BP+12} and \citet{Beaumont+12}.
This is illustrated in the left-lower panel of Fig.~\ref{fig:Npdf_dibujitos}, where we represent schematically the column density probability distribution function (\Npdf) of three different clouds {using lognormal functions}\footnote{For clarity, we used lognormal functions to illustrate this point. However, the explanation does not depend on the actual shape of the \Npdf, but on the fact that the \Npdf\ decreases rapidly.} (red, cyan and yellow dotted lines).  If we define the vertical double dot-dashed line in this figure as the column density threshold to define a cloud, we notice that, for each one of the three clouds, the mean column density above that threshold will be close to the threshold itself, since the few larger column density voxels will not contribute substantially to the average. This condition is satisfied in general terms by Galactic MCs: the \Npdf s of MCs decrease rapidly with column density, with exponents typically steeper than $-1.5$ \citep[see, e.g., ][]{Kainulainen+09}, regardless of whether the shape of the \Npdf\ is a lognormal, or power-law\footnote{It should also be stressed that, although there is some agreement that the \Npdf\ of MCs are either lognormal at low-column densities,  and power-law at larger, the  lognormality at low-column densities has been called into question by \citet[][]{Alves+17}, who argue that the \Npdf\ stops increasing when one goes from large to small column densities once one starts considering areas that are not fully sampled, i.e., for the incompleteness of maps, when accounting for non-closed contours in column density maps.\label{footnote:alves}}. Thus, the expected mass-size relation of a cloud ensemble defined by a single column density threshold and which have steep \Npdf s is precisely $M\propto R^2$.

Moreover, if we could measure volume densities and define them as connected voxels above some volume density, the result will be $M\propto R^3$ \citep[see ][]{BM02, BP+12} because the filling factor of the dense structures is small . In other words, the relation $M\propto R^2$ is real in the sense that it is an observational result not limited by the dynamical range of the observations \citep[e.g., ][]{Lombardi+10}. However, it does not represent any fundamental physical or structural state of clouds, but instead is the natural consequence of two facts: how we define an ensemble of clouds with a single column density threshold,  and the fact that the filling factor of the density decreases rapidly as density increases, which is translated into very steep \Npdf. 

In comparison with the extinction measurements, CO surveys exhibit larger scatter. But on top of that scatter, they exhibit systematically values of $q$ larger than 2, and thus, the natural question is why do larger clouds exhibit larger mean column densities than smaller clouds, when observed with CO? Should there be a single exponent? {Are we missing some fundamental physical property of the clouds related to CO emission?}

To answer these questions, we first note that an important difference between the extinction analysis by \citet{Lombardi+10} and the CO surveys as those from \citet{Roman-Duval+10} or \citet{Miville-Deschenes+17} is that the extinction analysis shows data for nearby MCs (basically, distances smaller than $\sim$600~pc), while the CO data {compiles cloud data}
from the whole Galaxy. The point is relevant because, as shown by \citet[][see their Fig.~24, upper panel]{Miville-Deschenes+17}, there is a clear observational bias: 
{as we observe more distant molecular clouds, we also observe larger clouds.}
This means that we are not able to split up far-away clouds into smaller clouds because of lack of resolution, as has been {noted} by \citet[][]{Miville-Deschenes+17}. But it also means that, either the lack of resolution, or the increase of distance, could be playing a role in increasing the mean density of clouds. 




At first glance, {the lower}
resolution for farther away clouds alone should not be the cause of a systematically increasing column density. Imagine again that we have the same three clouds as before, each one with its own \Npdf\ (see Fig.~\ref{fig:Npdf_dibujitos}, upper left panel, for a schematic view of the situation). Let us assume that each cloud has its own \Npdf\ (lower-left panel, red, cyan and orange histograms). Let us now put them at a distance such that we cannot distinguish them as individual entities, but as a single, big cloud (green contour).  Let us assume also that they do not overlap each other. If this large complex is resolved, the resulting \Npdf\ will be just the addition of the individual \Npdf s (green line in the lower-left panel of Fig.~\ref{fig:Npdf_dibujitos}). This addition does not move the resulting \Npdf\ horizontally, but vertically, and thus, the mean column density will also be close to the chosen column density threshold (dashed line).

Increasing the distance to the observed clouds may, instead, have another effect: it increases the probability of having 
overlap of different clouds in the same line of sight. Such overlap may increase  the mean column density. To show this effect, we now assume that our three clouds considered before overlap each other (see upper right panel in  Fig.~\ref{fig:Npdf_dibujitos}), in an arbitrary way. The resulting \Npdf\ is shifted towards column density values substantially larger than the original \Npdf s (purple line in the lower-right panel of Fig.~\ref{fig:Npdf_dibujitos}). If we adopt the same column density threshold used before to define clouds, we will notice that the cloud resulting from this superposition will have a substantially larger mean column density than the three individual clouds. In fact, for the values represented in the left panels of Fig.~\ref{fig:Npdf_dibujitos}, the three clouds, and the ``no-superposition" (green cloud) cloud will have a mean column density of 2$\times 10^{21}$ cm\alamenos 2, while the cloud resulting from the superposition (purple cloud in the right panels of Fig.~\ref{fig:Npdf_dibujitos}) will have a mean column density around 3.5$\times$ 10\ala{21} cm\alamenos 2, i.e., a mean column density close to the maximum of its \Npdf.

Even if we define clouds as CO peaks having a well-defined position-position-velocity values,
observations are not exempt of some degree of line-of-sight superposition. It can be expected, thus, that the farther away a cloud is, and/or the larger the cloud is, the more likely superposition can occur. If furthermore,  CO surveys  use constant intensity thresholds to find clouds \citep[e.g., ][]{Roman-Duval+10, Miville-Deschenes+17}, it is likely that the situation illustrated in the right panels of Fig.~\ref{fig:Npdf_dibujitos} may be occurring. In the next section we explore this possibility, and quantify whether superposition of CO emission in the line of sight can account for increasing the exponent in the mass-size relation from $q=2$ to $q\sim 2.2-2.4$.

\section{Superposition of clouds when using CO surveys}\label{sec:model}

Let us consider that, on average, the extinction of the interstellar medium of our Galaxy increases at a rate\footnote{We checked the validity of this value by using the average extinction {\it vs.} distance tabulation of \citet{Chen+13} from which we obtained a linear increase of $\sim 1.5$ mag kpc\alamenos 1.} of $\sim$1.6 mag kpc\alamenos 1 \citep[e.g., ][]{BinneyMerrifield98}, i.e., 

\begin{equation}
    \frac{dN}{dx} \sim 10^{21} {\rm cm^{-2}\  kpc^{-1} }
    \label{eq:mean_extinction}
\end{equation}
(with $x$ the distance in the line of sight), which corresponds to a mass column density variation of

\begin{equation}
    \frac{d\Sigma}{dx} \sim 10-20\ M_\odot\ {\rm pc}^{-2}\  {\rm kpc}^{-1},
    \label{eq:mean_mass_columndensity}
\end{equation}
depending on whether the column is mostly atomic, with mean atomic weight of $\mu = 1.27$, or molecular ($\mu = 2.36$, using Solar abundances in both cases). When observing molecular clouds in CO emission, smaller clouds in front or behind the main cloud, at similar velocity, will not be distinguished from the big cloud. In fact, this effect could be actually quite important if our Milky Way has substantial amount of small clouds in the inter-arm region, as has been estimated by \citet{Koda+16}.

It is evident that only a fraction of the mass in the line of sight will be in the form of CO, and also only a fraction of that CO will be at the right velocities. Thus, assuming eq.~(\ref{eq:mean_mass_columndensity}) as a reasonable average in the Galaxy, the total mass computed for the observed cloud should be 

\begin{equation}
    M = \biggl(\Sigmath + f\ x\ \frac{ d \Sigma } {dx}  \biggr) R^2
    \label{eq:M-R_mod}
\end{equation}
{where the first term within parenthesis is the mass surface density of the cloud if there were no LOS superposition, and the second term is the fraction $f$ of the actual mass surface density $x d\Sigma/dx$ that is in the same line of sight,} 
that is in molecular form and has the same velocity of the cloud, and thus, that will be computed as part of the cloud.  

As discussed by \citet[][]{Miville-Deschenes+17}, the finite angular resolution and sensitivity of the data translates into a linear size-distance correlation, which we write as:

\begin{equation}
    R \simeq \gamma\ x .
    \label{eq:d-R}
\end{equation}
{where $\gamma$ is a scale factor to be derived from the observational data (see below).} This allows us to rewrite eq. (\ref{eq:M-R_mod}) as:

\begin{equation}
    M = \biggl(\Sigmath + \frac{f}{\gamma}\ \frac{ d \Sigma } {dx}\  R \biggr) R^2
    \label{eq:M-R_mod2}
\end{equation}

{Following  \citet{Miville-Deschenes+17}, we can estimate either $\Sigmath$ and $\gamma$. In the first case, the authors use a threshold of $W_{\rm CO} = 0.8$~K~km~sec\alamenos 1 (see their section 2.4.2). This is translated into a column density of $\Sigmath \sim 3.5$\Msun\ pc$^{-2}$.  In the second case, we estimate $\gamma$ from their Fig.~24a, by 
noticing that the bisectrix\footnote{Since these data points are not available online, we did not computed a formal fit.} to the
$R-D$ data goes through a point at ($D=1$~kpc,  $R\sim 4.7$~pc), with a logarithmic scatter of 0.278 dex. Thus, we assume $\gamma \sim 4.7$ pc/kpc. 
}

\begin{figure}
    \includegraphics[width=\columnwidth]{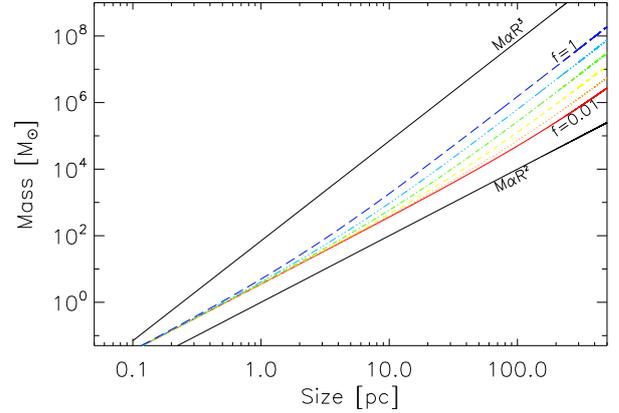}
    \caption{Mass-size relation from eq.~(\ref{eq:M-R_mod2}). From bottom to top, we show 6 models with $f \in (0.01,1)$, equally spaced in logarithm of $f$. Additionally, for reference we show with black solid lines the mass-size relation with exponents $q=$ 2 and 3. (See electronic version for a color version of this figure).}
    \label{fig:MR_noscatter}
\end{figure}

Taking $d\Sigma/dx$ from eq. (\ref{eq:mean_mass_columndensity}), we plot in Fig.~\ref{fig:MR_noscatter} the mass-size relation as given by eq.~(\ref{eq:M-R_mod2}) in the ranges $R \in (5\times 10^{-2}, 500)$~pc, $M\in (10^{-1}, 10^9)$~\Msun, substantially larger than the typical ranges for which the mass-size relation is typically reported for Galactic studies. It is clear from this plot that, for the fiducial parameters in eq.~(\ref{eq:M-R_mod2}), the models show a transition between the slope of 2 and 3 at scales between a fraction of a parsec and $\sim$100~pc, the typical scales of MCs. It is straightforward to calculate the transition radius between the quadratic and cubic behavior of eq.~(\ref{eq:M-R_mod2}) is given by:

\begin{equation}
    \Rtrans \sim  \frac{\gamma}{f}
    \frac{\Sigmath}{d\Sigma/dx} = 1.65\ \frac{1}{f}\ {\rm pc}
    \label{eq:Rtrans}
\end{equation}
{This transition radius means that, for values of $f\in (10^{-3},1)$, the transition between the quadratic and cubic behaviour occurs at scales of MC or even GMCs. However, one should not expect $f$ to be unique for all clouds}.

\begin{figure}
    \includegraphics[width = \columnwidth]{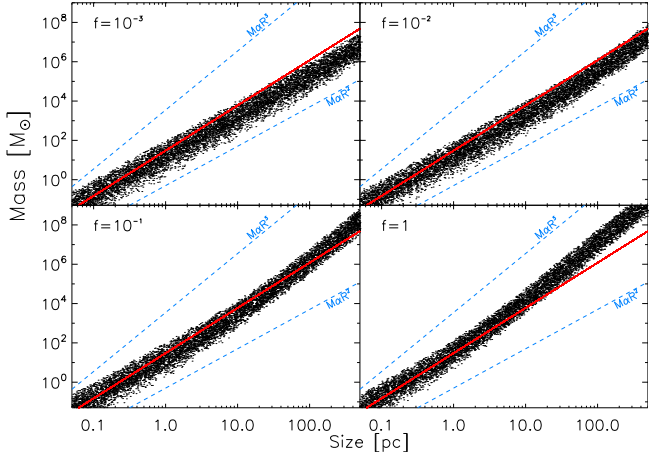}
    \caption{Mass-size relation according to eq.~(\ref{eq:M-R_mod2}), using the value of $f$ as indicated in each panel, and spreading out the mass by a factor of 10. The transition between the quadratic and cubic regime is still noticeable.}
     
    \label{fig:MR_scatter}
\end{figure}

In Fig.~\ref{fig:MR_scatter} we present models with $f=0.001$ (upper left panel), $f=0.01$ (upper right), $f=0.1$ (lower left) and $f=1$ (lower right), assuming the fiducial values for eq.~(\ref{eq:M-R_mod2}), but spreading out randomly the obtained mass within a factor of 10. This allow us to account for possible uncertainties in the determination of mass and the distance to the cloud (and thus, its actual size). For reference, we show the power-laws with exponents of 2 and 3 (blue dashed lines) as well as a power-law with exponent 2.3 (red solid line), the value frequently quoted in the literature for the mass-size relation. 

Although at first glance Fig.~\ref{fig:MR_scatter} reproduces qualitatively the behavior of the observational data, we can clearly notice that 
the trends are curved,
{according to  eq.~(\ref{eq:Rtrans}): for $f=1$, $10^{-1}$, $10^{-2}$ and $10^{-3}$,  $\Rtrans \sim 1.65$~pc, 16.5~pc, 165~pc and 1,650~pc, respectively. } 

\begin{figure}
    \includegraphics[width=\columnwidth]{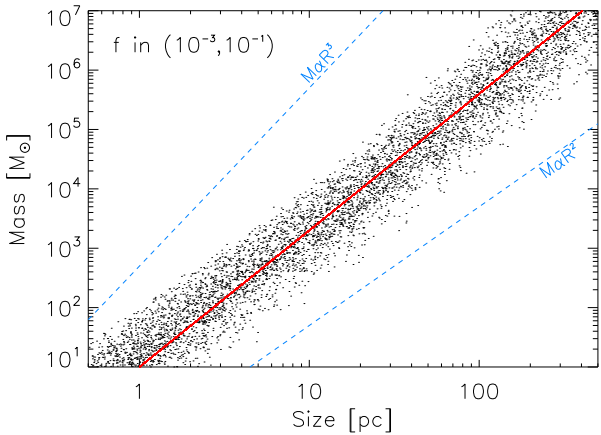}
    \caption{Mass-size relation for eq.~(\ref{eq:M-R_mod2}) using random values of $f$ between $10^{-3}$ and $10^{-1}$ chosen uniformly in log-space, and spreading out the mass by a factor of 10. We also have limited the dynamical range to that of \citet{Miville-Deschenes+17}. 
    The transition between the quadratic and cubic regime has disappeared, and the tendency of the data exhibits a slope of $q\sim 2.36$.}
    \label{fig:MR_scaterOne}
\end{figure}

It should be noted, however, that such curved feature does not appear in the published mass-size relations. In order to sort this concern out, two 
aspects
should be noticed. First
$f$ represents the amount of additional mass that one will compute that does not belong to a
cloud, but it is assigned to it because it is at the 
same
velocity. This value is fixed in each panel of Fig.~\ref{fig:MR_scatter}, but there is no reason for it to be so. 
Second,
it should also be noticed that the range of masses and sizes in Fig.~\ref{fig:MR_noscatter} and \ref{fig:MR_scatter} is actually larger than the range of sizes and masses in CO surveys \citep[e.g., ][]{Roman-Duval+10, Miville-Deschenes+17}. Consequently, in Fig.~\ref{fig:MR_scaterOne}, we plot the mass-size relation, where we have spread out uniformly (in $\log{f}$) the value of $f$, in the range $f\in (10^{-3}, 10^{-1})$, and we have limited the mass and size ranges to the values presented in the literature, i.e.,  $M \in (10,10^7)~\Msunmath$ and $R\in (0.05,500)$~pc \citep[][]{Miville-Deschenes+17}. As before, the blue dashed lines represent the mass-size relations with exponents $q=2$, 3, and the red solid line $q=2.3$. It becomes clear now that the trend and scatter are similar to that found in CO surveys.

\section{Discussion and implications}\label{sec:discussion}

{Before discussing the implications and extensions of our result, it should be noted that, since the introduction, we have quoted systematically \citet{Roman-Duval+10} and \citet{Miville-Deschenes+17} as the works on reporting the mass-size relationship for MCs on Galactic scale, reporting a power-law exponent of $q>2$, specifically, $q=2.2-2.36$. It is important to note, however, that although it is not shown a mass-size relation, the work by \citet{Solomon+87} also studied clouds on Galactic scale. This is quoted as one of the first studies that reports a very clear velocity dispersion-size power-law  relation  ($\sigma_v\propto R^{1/2}$), and in which clouds are in near virial equilibrium. Implicitly, thus, it has been frequently assumed that the clouds in this survey exhibit nearly constant column density \citep[see, e.g., ][]{Scalo90, Heyer+09, Faesi+18}, and thus, a mass-size relationship of with $q\sim2$. However, considering that the CO luminosity is proportional to the luminous mass of the cloud, when the CO luminosity-size relationship is plotted for the \citet{Solomon+87} data, a clear mass-size power-law relationship emerges, with a slope of $\sim 2.5$  \citep[see, e.g., Fig. 3 in ][]{Bolatto+08}.  Thus, strictly speaking, the work by \citet{Solomon+87} has no constant column density, but instead, is consistent with the works by \citet{Roman-Duval+10} and \citet{Miville-Deschenes+17}, showing $q>2$. As a corollary, let us say finally that, rather than a constant virial parameter, the results from \citet{Solomon+87} imply that $\alphavir \propto M^{-0.2}$, or equivalently, $\alphavir\propto R^{-1/2}$, more consistently with typical estimations of the energy budget of clouds \citep[e.g., ][]{BertoldiMcKee92, Kauffmann+13}.
}

The model presented in the previous sections indicates that the mass-size relationships with slope steeper than 2 for MCs observed in CO through the whole Galaxy are likely the result of superposition of clouds in the line of sight. 

One of the hypothesis used to obtain this result is that there is a fraction $f$ of molecular gas in the line of sight of the cloud that does not correspond to the cloud itself, but that will contribute to the emission at the velocities of the cloud. We found that a varying factor between 0.01\%\ and 10\%\ of the mass along the line of sight may explain the value of $q\sim 2.36$ (see Fig.~\ref{fig:MR_scaterOne}).

{
Mass-size power-law relationships with exponents larger than 2 imply that larger GMCs should have larger mean column densities. For the particular case of the observations of clouds in the Milky Way, this means that farther away clouds should have larger mean column densities. This could be a reasonable result if we naively consider that the properties of GMCs may depend on their position in their host galaxy, as found by \citet{Colombo+14}. However, such dependency is not arbitrary, but it is intimately related to the inner structure of the host galaxy, not on the distance to a particular point on the galaxy, as it is the case of the observations of clouds in the Milky Way.  
As commented before, Galactic clouds show a dependency of the column density with the distance to the Sun, regardless the particular environment in which the cloud could be. 
A similar result is found by \citet{Roman-Duval+10}. And 
since there is no physical reason for which MCs at increasingly distances from the Sun should have increasingly larger column densities, superposition effects seem a plausible explanation. } 


Our explanation for the mass-size relations found with slopes larger than 2 still faces a challenge: different studies of (a) cores in single clouds, (b) trunks, branches and leaves within a single cloud, and (c) clouds in a single external galaxy, does not necessarily exhibit a mass-size relationship with $q\sim 2$, though they are basically at the same distance from us. In the next sections we will discuss these issues.

\subsection{Mass-size relation for dense cores withing single clouds}

{
Different studies of the structure  of a single molecular clouds in our Galaxy shows that their cores do not necessarily exhibit a mass-size relation with exponent $q=2$ \citep[e.g., ][]{Williams+94, Loren89, ElmegreenFalgarone96, Johnstone+00, Lada+08, RomanZuniga+10, Kainulainen+11, Konyves+15, Kirk+17,  Veltchev+18}. Belonging to the same cloud, the argument that the farther away, the larger probability of superposition in the LOS does not work. 
}


There are several reasons why this variety of results can be obtained for dense cores within a single cloud: 

\begin{enumerate}

  \item When in some way, a single volume density threshold can be used to find cores, the slope of the mass-size relation is closer to $q=3$ \citep[e.g., ][]{Kainulainen+11}.

  \item When multi-threshold methods are used, but still the reported mass-size relation is that of the cores without inner substructure, there is not necessarily a single slope, as it may be the case of the algorithms getsources, gaussclumps, or the leaves in dendrograms \citep[e.g., ][]{Konyves+15, Kirk+17, Veltchev+18, Sanhueza+19}.
  
  \item {While the expected value of $q$ for clumpfinding methods that use column density thresholds is 2 \citep{Lombardi+10, BP+12, Beaumont+12}, note that even for this method in some cases the slope found is larger \citep[e.g., ][]{Williams+94, Johnstone+00, Lada+08, RomanZuniga+10}. In principle, larger powers for the mass-size power-law relation could still be the result of superposition, as theoretical and observational work has suggested. In the former group, for instance,  \citet{BM02},  \citet{Gammie+03} and \citet{Shetty+10} analyse the velocity structure of cores in simulations of MCs, finding that indeed, the velocity structure of what will be observed as cores is the result of the actual velocity structure of the physical core, plus some superpositon of line of sight material that does not belong to the core itself. Even the fibers, which are carefully defined as elongated position- and velocity-coherent structures \citep{Hacar+13}, seem to be the result of line of sight superposition \citep{ZA+17, Clarke+18}. And the larger the core in a cloud appears in observational data, the larger the probability that its emission is the result of line of sight superposition of material in the line of sight. If this occurs, then it will not be strange that the mass-size relation has a slope larger than $q>2$.}
{From the observational perspective, in their study of the Pipe Nebula, \citet{Frau+15} used a $^{13}$CO map by \citet{Onishi+99} to present evidence 
of some degree of overlap along the line of sight for some cores. }

\end{enumerate}

In summary, there are several possibilities that could give rise to a larger slope in the mass-size relation for cores in a single cloud, including line of sight superposition.  A detailed study on the different clumpfinding methods used, and slopes obtained, will be presented elsewhere (Román-Zúñiga \& Ballesteros-Paredes 2019, in preparation). 

{A final word of caution should be given: it is important to clarify that what we intend to discuss in the present subsection is the mass-size relationship for multiple cores in a single cloud. It should not be confused with  the single-cloud mass-size relationship \citep[e.g., ][]{Lombardi+10, Kauffmann+10a, Kauffmann+10b}, where the inner substructures at different levels of the hierarchy are not necessarily dense cores. In this case, the slope of the mass-size relationship is necessarily $q \le 2$  and it depends directly on the slope of the column density PDF of the cloud, as has been discussed by \citet{BP+12}}.

\subsection{The extragalactic mass-size relation}

{
Clouds in a single external galaxy are also basically at the same distance from us, and thus, they lack the distance effect previously discussed. If such clouds are thresholded at the same column density, we expect them to exhibit a mass-size power-law relation with exponent $q\sim 2$ (provided that their column density PDF decays rapidly). This result is indeed, found by \citet{Faesi+18} and \citet{Hughes+13}. The first authors analyze in great detail molecular clouds in the galaxy NGC 300 at resolution of $\sim$~10~pc. The second one made a detailed comparison between the data set of three galaxies, M51, M33 and the LMC. For each of these galaxies, these authors argue that, due to limited resolution and sensitivity, their correlations between luminosity and size can be an artifact. In Hughe's et al. words, ``GMCs in each panel typically lie close to the surface brightness sensitivity limits of each survey", supporting the idea that clouds, when chosen by sensitivity thresholds, will appear  to have similar column densities \citep{BM02, BP+12, Beaumont+12}. 
}

{The situation for extragalactic studies, however, is not that straightforward. For instance, \citet{Bolatto+08} reports a slope of $\sim$~2.5 for their sample, while
the  work by \citet{Hughes+13} quoted above exhibits  substantial scatter, depending on how the clouds are considered, if as large complexes, or decomposed into individual structures. Such scatter for each individual galaxy resembles somehow the case of M51 alone studied by \citet{Colombo+14}: when all clouds of M51 are included, there is a poor correlation coefficient.}

\citet{Colombo+14}, furthermore, found important differences between the variety of environments in the same galaxy. For the spiral arms, these authors found a mass-size relation with $q\sim 2.4$, for the inner galaxy, $q\sim 2$, and for the inter-arm region, even flatter, $q\sim 1.5$. The origin of such differences, as the authors acknowledge, is likely to be the different CO emission properties within the different M51 environments, as well as geometry effects and filling factors. The clouds in the extragalactic studies have resolutions of $> 30$~pc \citep[only][reaches resolutions of $\sim$~10~pc]{Faesi+18}, while the local GMCs in the Milky Way are already of such sizes (10--50~pc), and thus, superposition effects between separated clouds are likely ocurring. Thus, better studies with improved resolution are needed to understand the nature of the scaling in extragalactic studies, and compare it with the results of the Milky Way.

A crucial point that remains to be understood is the fact that, on average, GMC properties (mass, size, luminosity, velocity dispersion) in a set of local galaxies are similar than those properties of Galactic clouds \citep{Bolatto+08}. In particular, this implies that the mass-size relationship for those galaxies has a power-law exponent of $q\sim 2.5$, since the comparison that \citet{Bolatto+08} made is with \citet[][see \S\ref{sec:discussion} above]{Solomon+87}. Such a result is apparently in disagreement with the results by \citet{Hughes+13}, \citet{Faesi+18} ($q \sim 2$) and \citet{Colombo+14} (poor correlations, dependency on the environment) quoted above.  A close inspection of \citet{Bolatto+08}'s Fig.~3 shows, however, that only the LMC exhibits clearly a slope consistent with $q\sim 2.5$, but that, individually, the few data points for every other galaxy in the \citet{Bolatto+08}'s sample, exhibit a quite flat mass-size relationship. What appears to be occurring in this case is that, individually, there are not enough statistics in the \citet{Bolatto+08} sample to construct a mass-size relationship for each galaxy, as could be the case of \citet{Hughes+13} and  \citet{Colombo+14}. However, when \citet{Bolatto+08} consider the mass-size relationship for the whole set of galaxies together, the mean column density of the clouds in larger galaxies tends to be larger. This is consistent with the scenario where clouds in different galaxies are also under different galactic environments (galactic potential, galactic pressure, etc), in agreement with the result by \citet{Colombo+14}.

{
We cannot discard the idea that different environments can produce clouds with different mean column densities. For instance, a deeper potential might produce larger column density clouds.
However, what is important for our discussion is that one can also expect that different environments could be more prone to superposition effects. For instance, clouds in the arms could have substantially more overlapping effects than clouds in the inter-arm regions, explaining thus the differences in the mass-size relationships for different regions found by \citet{Colombo+14}. Their results, furthermore, impose a cautionary warning for the study of the mass-size relationship of our own Galaxy: if we are unable to distinguish whether an observed Galactic cloud is in one or another environment, we might be producing a mass-size relationship that arises from clouds in  different physical environments. 
}

It is convenient to stress that, although \citet{Colombo+14} made an effort to avoid line blending in their cloud decomposition, the situation can be quite complex, since even at the smallest scales, where coherent structures can be found (i.e., cores, fibers), substantial superposition along the line of sight can occur \citep{BM02, Gammie+03, Shetty+10, ZA+17, Clarke+18}.

{
The situation for clouds in external galaxies might deserve a full further study, since the resolution is substantially different in external Galaxies \citep[typically, 30--50~pc, see, e.g., ][]{Utomo+18, Sun+18}, with only one study \citep{Faesi+18} achieving resolutions of $\sim$~10~pc. A first puzzling challenge is to understand, for instance, the origin of  mass-size relation with $q<2$ in the inter-arm region of M51, since $q=2$ is the limit when clouds are defined via thresholds in column density.  
}

\subsection{Fractal dimension?}

Different studies have interpreted the exponent of the power-law mass-size relation as representative of the fractal dimension \citep[e.g., ][]{ElmegreenFalgarone96, Roman-Duval+10, Beattie+19}. This approach has two hypotheses: first, that the mass can be a proxy for the area. Second, that the linear size is related also to the area in a quadratic way, since in the observational studies, the size is  typically considered to be proportional to the square root of the area of the cloud. 

In principle, the first hypothesis seems plausible:
observed molecular clouds exhibit 
{rapidly decaying }
\Npdf s \citep{Kainulainen+09}. The same is found for molecular clouds in numerical simulations \citep[e.g., ][]{VSGarcia01, BP+11b, Kritsuk+11, Tassis+10}. This means that the filling factor of the dense structures is quite small, which in turn, implies that the mass should be proportional to the area. 

{
However, for a collection of Galactic clouds, the superposition effects we discussed are not compatible with invoking mass as a proxy for the area. As shown in the present work, the same fact that the exponent in the mass-size relation is larger than 2 means that the mass is not proportional to the area. In addition
}
invoking the square root of the area as a proxy for the linear size does not tells us what will be the linear size-area correlation, which is precisely what is intended to be measured with the fractal dimension. 
A more appropriate estimation of the fractal dimension will be that of comparing the actual area of an ensemble of clouds, with an independent estimation of their linear size, e.g., its maximum extension.  {Surprisingly, there is no work in the field of molecular clouds for which the fractal dimension has been calculated by computing the area, rather than mass, and a linear size measurement independent to the area. }

Finally, it is interesting to note that the expected value for the fractal dimension of clouds is between 2 and 3. Therefore, the slope of the  mass-size relation for CO clouds and cores, which is likely contaminated by superposition effects, matches somehow the expected value for the fractal dimension of molecular clouds, though it is not the actual fractal dimension. This is probably the reason the estimations of the fractal dimension {\it via} the mass-size relation have been assumed to be correct.



\subsection{The meaning of the mass size relationship}

As we have discussed throughout the manuscript, once the density PDF of molecular clouds is steep enough, the resulting slope of the mass-size relation for an ensemble of clouds depends directly on the definition of clouds. If column density thresholds are used and no superposition effects are present, the slope is around 2. If instead, larger clouds exhibit larger superposition effects, one can expect slopes larger than 2. Furthermore, if volume density thresholds are used, the slope is 3. If instead, different thresholds are implemented, there is not unique answer. If larger superposition effects occur for bigger regions, the slope can be larger than 2, even if column density thresholds are used. 
Therefore, finding a mass-size relationship with $q\sim 2$, does not mean that the result has implications on the dynamical state of molecular clouds. Instead, the result depends primarily on how clouds are defined. Moreover, a slope of 2 does not mean that the column density of the sample is constant. In fact, it is interesting to notice that the data shown by \citet[][]{Faesi+18} has a spread over more than one order of magnitude in column density (see their Fig.~13), although the authors argue that the column density is constant.

Finally, the common argument in the literature that clouds following
Larson relations with exponents $1/2$ and $2$ for the velocity dispersion-size and mass-size relations, respectively,  
implies that clouds are in virial equilibrium, seems not to be appropriate. In order to determine whether a single cloud is actually in virial equilibrium or not, should require a careful analysis of the cloud itself. In fact, by using different thresholds, the mass-size relationships of a single cloud has necessarily a power-index smaller than 2 \citep[][see Ballesteros-Paredes et al. 2012 for an explanation]{Kauffmann+10a, Kauffmann+10b, Lombardi+10}. If one thinks that the velocity-size relationship with power-index of 1/2 holds for that cloud, then it will not be in virial equilibrium.

\section{Conclusions}\label{sec:conclusion}

{
In the present contribution we have discussed the mass-size relation, $M\propto r^q$, for an ensemble of molecular clouds. We show that the difference between the extinction studies for nearby clouds, which show a tight power-law correlation with exponent $q=2$, and studies of Galactic clouds observed in CO,  showing $2 < q < 3$ with a substantial scatter, is due to superposition of CO emission in the line of sight of far away clouds. 
}

{
We arrive to this conclusion by noticing that the farther away the clouds are, the larger their mean column densities. Since the physics of the clouds all around the Galaxy should be, on average, similar, there is not an intrinsic (i.e., physical) reason why distant molecular clouds should have larger column densities. We have estimated that a random fraction of gas in the line of sight oscillating between 0.001\%\ and 10\%\ can explain the exponent of $q\sim 2.3$ found in observational Galactic surveys.
}

{
In the case of dense cores within a single cloud, we suggest a couple of possibilities that could be playing a role:
First of all, the method and threshold used to find clouds could play a role in the obtained mass-size relation. Second, superposition effects are still more likely to occur for larger cores than for smaller cores, artificially increasing the mass of the observed core. 
}

{
For the extragalactic case, we noticed that still, some results appear to be consistent with the idea of the threshold defining a mass-size relationship with slope of 2. However, there is strong scatter in the data, and in reality, it is hard to argue in favour of a single mass-size relationship. Environmental differences can play a role in changing the slope of the mass-size relationship \citep[e.g., ][]{Colombo+14}, a result consistent with our idea of superposition as the responsible for increasing the slope: clouds in the arms, with larger slopes in their mass-size relationship, should be more prone to line-of-sight superposition than clouds in the inter-arm region, with flatter mass-size relationships.   }

We have argued also that the mass-size relation should NOT be used as equivalent to the fractal dimension, since the mass is not proportional to the area, and the typical size used in those relations is  a linear size independent of the area.
Finally, we have argued that the statistical analysis of finding clouds satisfying the Larson's relations does not mean that each individual cloud is in virial equilibrium. In order to understand the dynamical state of a cloud it is necessary to understand in detail the actual structure of every single cloud.

\section*{Acknowledgments}

J.B-P and CRZ acknowledge  UNAM-DGAPA-PAPIIT support through grant numbers IN-111-219 and IN108-117, respectively. Q.S. acknowledges postdoctoral fellowship from UNAM-DGAPA 10653. M.Z.A. acknowledges support from CONACYT grant number A1-S-54450 to Abraham Luna Castellanos (INAOE). MJJD  acknowledges support from the Smithsonian Institution as a Sub-millimeter  Array  (SMA)  Fellow. This work has made extensive use of the NASA-ADS database.

\bibliographystyle{mnras}
\bibliography{references} 

\begin{thebibliography}{}
\makeatletter
\relax
\def\mn@urlcharsother{\let\do\@makeother \do\$\do\&\do\#\do\^\do\_\do\%\do\~}
\def\mn@doi{\begingroup\mn@urlcharsother \@ifnextchar [ {\mn@doi@}
  {\mn@doi@[]}}
\def\mn@doi@[#1]#2{\def\@tempa{#1}\ifx\@tempa\@empty \href
  {http://dx.doi.org/#2} {doi:#2}\else \href {http://dx.doi.org/#2} {#1}\fi
  \endgroup}
\def\mn@eprint#1#2{\mn@eprint@#1:#2::\@nil}
\def\mn@eprint@arXiv#1{\href {http://arxiv.org/abs/#1} {{\tt arXiv:#1}}}
\def\mn@eprint@dblp#1{\href {http://dblp.uni-trier.de/rec/bibtex/#1.xml}
  {dblp:#1}}
\def\mn@eprint@#1:#2:#3:#4\@nil{\def\@tempa {#1}\def\@tempb {#2}\def\@tempc
  {#3}\ifx \@tempc \@empty \let \@tempc \@tempb \let \@tempb \@tempa \fi \ifx
  \@tempb \@empty \def\@tempb {arXiv}\fi \@ifundefined
  {mn@eprint@\@tempb}{\@tempb:\@tempc}{\expandafter \expandafter \csname
  mn@eprint@\@tempb\endcsname \expandafter{\@tempc}}}

\bibitem[\protect\citeauthoryear{{Alves}, {Lombardi}  \& {Lada}}{{Alves}
  et~al.}{2017}]{Alves+17}
{Alves} J.,  {Lombardi} M.,   {Lada} C.~J.,  2017, \aap, 606, L2

\bibitem[\protect\citeauthoryear{{Ballesteros-Paredes}}{{Ballesteros-Paredes}}{2006}]{BP06}
{Ballesteros-Paredes} J.,  2006, \mn@doi [\mnras]
  {10.1111/j.1365-2966.2006.10880.x}, \href
  {https://ui.adsabs.harvard.edu/abs/2006MNRAS.372..443B} {372, 443}

\bibitem[\protect\citeauthoryear{{Ballesteros-Paredes} \& {Mac
  Low}}{{Ballesteros-Paredes} \& {Mac Low}}{2002}]{BM02}
{Ballesteros-Paredes} J.,  {Mac Low} M.-M.,  2002, \apj, 570, 734

\bibitem[\protect\citeauthoryear{{Ballesteros-Paredes}, {Klessen}, {Mac Low}
  \& {Vazquez-Semadeni}}{{Ballesteros-Paredes} et~al.}{2007}]{BP+07}
{Ballesteros-Paredes} J.,  {Klessen} R.~S.,  {Mac Low} M.~M.,
  {Vazquez-Semadeni} E.,  2007, in {Reipurth} B.,  {Jewitt} D.,   {Keil} K.,
  eds, Protostars and Planets V. p.~63

\bibitem[\protect\citeauthoryear{{Ballesteros-Paredes}, {Hartmann},
  {V{\'a}zquez-Semadeni}, {Heitsch}  \&
  {Zamora-Avil{\'e}s}}{{Ballesteros-Paredes} et~al.}{2011a}]{BP+11a}
{Ballesteros-Paredes} J.,  {Hartmann} L.~W.,  {V{\'a}zquez-Semadeni} E.,
  {Heitsch} F.,   {Zamora-Avil{\'e}s} M.~A.,  2011a, \mnras, 411, 65

\bibitem[\protect\citeauthoryear{{Ballesteros-Paredes}, {V{\'a}zquez-Semadeni},
  {Gazol}, {Hartmann}, {Heitsch}  \& {Col{\'\i}n}}{{Ballesteros-Paredes}
  et~al.}{2011b}]{BP+11b}
{Ballesteros-Paredes} J.,  {V{\'a}zquez-Semadeni} E.,  {Gazol} A.,  {Hartmann}
  L.~W.,  {Heitsch} F.,   {Col{\'\i}n} P.,  2011b, \mnras, 416, 1436

\bibitem[\protect\citeauthoryear{{Ballesteros-Paredes}, {D'Alessio}  \&
  {Hartmann}}{{Ballesteros-Paredes} et~al.}{2012}]{BP+12}
{Ballesteros-Paredes} J.,  {D'Alessio} P.,   {Hartmann} L.,  2012, \mnras, 427,
  2562

\bibitem[\protect\citeauthoryear{{Ballesteros-Paredes}, {V{\'a}zquez-Semadeni},
  {Palau}  \& {Klessen}}{{Ballesteros-Paredes} et~al.}{2018}]{BP+18}
{Ballesteros-Paredes} J.,  {V{\'a}zquez-Semadeni} E.,  {Palau} A.,   {Klessen}
  R.~S.,  2018, \mnras, 479, 2112

\bibitem[\protect\citeauthoryear{{Beattie}, {Federrath}  \&
  {Klessen}}{{Beattie} et~al.}{2019}]{Beattie+19}
{Beattie} J.~R.,  {Federrath} C.,   {Klessen} R.~S.,  2019, \mnras, 487, 2070

\bibitem[\protect\citeauthoryear{{Beaumont}, {Goodman}, {Alves}, {Lombardi},
  {Rom{\'a}n-Z{\'u}{\~n}iga}, {Kauffmann}  \& {Lada}}{{Beaumont}
  et~al.}{2012}]{Beaumont+12}
{Beaumont} C.~N.,  {Goodman} A.~A.,  {Alves} J.~F.,  {Lombardi} M.,
  {Rom{\'a}n-Z{\'u}{\~n}iga} C.~G.,  {Kauffmann} J.,   {Lada} C.~J.,  2012,
  \mn@doi [\mnras] {10.1111/j.1365-2966.2012.21061.x}, 423, 2579

\bibitem[\protect\citeauthoryear{{Bertoldi} \& {McKee}}{{Bertoldi} \&
  {McKee}}{1992}]{BertoldiMcKee92}
{Bertoldi} F.,  {McKee} C.~F.,  1992, \apj, 395, 140

\bibitem[\protect\citeauthoryear{{Binney} \& {Merrifield}}{{Binney} \&
  {Merrifield}}{1998}]{BinneyMerrifield98}
{Binney} J.,  {Merrifield} M.,  1998, {Galactic Astronomy}

\bibitem[\protect\citeauthoryear{{Bolatto}, {Leroy}, {Rosolowsky}, {Walter}  \&
  {Blitz}}{{Bolatto} et~al.}{2008}]{Bolatto+08}
{Bolatto} A.~D.,  {Leroy} A.~K.,  {Rosolowsky} E.,  {Walter} F.,   {Blitz} L.,
  2008, \mn@doi [\apj] {10.1086/591513}, 686, 948

\bibitem[\protect\citeauthoryear{{Chen}, {Schultheis}, {Jiang}, {Gonzalez},
  {Robin}, {Rejkuba}  \& {Minniti}}{{Chen} et~al.}{2013}]{Chen+13}
{Chen} B.~Q.,  {Schultheis} M.,  {Jiang} B.~W.,  {Gonzalez} O.~A.,  {Robin}
  A.~C.,  {Rejkuba} M.,   {Minniti} D.,  2013, \mn@doi [\aap]
  {10.1051/0004-6361/201219682}, 550, A42

\bibitem[\protect\citeauthoryear{{Clarke}, {Whitworth}, {Spowage},
  {Duarte-Cabral}, {Suri}, {Jaffa}, {Walch}  \& {Clark}}{{Clarke}
  et~al.}{2018}]{Clarke+18}
{Clarke} S.~D.,  {Whitworth} A.~P.,  {Spowage} R.~L.,  {Duarte-Cabral} A.,
  {Suri} S.~T.,  {Jaffa} S.~E.,  {Walch} S.,   {Clark} P.~C.,  2018, \mnras,
  479, 1722

\bibitem[\protect\citeauthoryear{{Colombo} et~al.,}{{Colombo}
  et~al.}{2014}]{Colombo+14}
{Colombo} D.,  et~al., 2014, \mn@doi [\apj] {10.1088/0004-637X/784/1/3}, 784, 3

\bibitem[\protect\citeauthoryear{{Dame}, {Hartmann}  \& {Thaddeus}}{{Dame}
  et~al.}{2001}]{Dame+01}
{Dame} T.~M.,  {Hartmann} D.,   {Thaddeus} P.,  2001, \apj, 547, 792

\bibitem[\protect\citeauthoryear{{Elmegreen} \& {Falgarone}}{{Elmegreen} \&
  {Falgarone}}{1996}]{ElmegreenFalgarone96}
{Elmegreen} B.~G.,  {Falgarone} E.,  1996, \apj, 471, 816

\bibitem[\protect\citeauthoryear{{Faesi}, {Lada}  \& {Forbrich}}{{Faesi}
  et~al.}{2018}]{Faesi+18}
{Faesi} C.~M.,  {Lada} C.~J.,   {Forbrich} J.,  2018, \mn@doi [\apj]
  {10.3847/1538-4357/aaad60}, 857, 19

\bibitem[\protect\citeauthoryear{{Frau}, {Girart}, {Alves}, {Franco}, {Onishi}
  \& {Rom{\'a}n-Z{\'u}{\~n}iga}}{{Frau} et~al.}{2015}]{Frau+15}
{Frau} P.,  {Girart} J.~M.,  {Alves} F.~O.,  {Franco} G.~A.~P.,  {Onishi} T.,
  {Rom{\'a}n-Z{\'u}{\~n}iga} C.~G.,  2015, \mn@doi [\aap]
  {10.1051/0004-6361/201425234}, 574, L6

\bibitem[\protect\citeauthoryear{{Gammie}, {Lin}, {Stone}  \&
  {Ostriker}}{{Gammie} et~al.}{2003}]{Gammie+03}
{Gammie} C.~F.,  {Lin} Y.-T.,  {Stone} J.~M.,   {Ostriker} E.~C.,  2003, \apj,
  592, 203

\bibitem[\protect\citeauthoryear{{Hacar}, {Tafalla}, {Kauffmann}  \&
  {Kov{\'a}cs}}{{Hacar} et~al.}{2013}]{Hacar+13}
{Hacar} A.,  {Tafalla} M.,  {Kauffmann} J.,   {Kov{\'a}cs} A.,  2013, \mn@doi
  [\aap] {10.1051/0004-6361/201220090}, 554, A55

\bibitem[\protect\citeauthoryear{{Hartmann}, {Ballesteros-Paredes}  \&
  {Bergin}}{{Hartmann} et~al.}{2001}]{HBB01}
{Hartmann} L.,  {Ballesteros-Paredes} J.,   {Bergin} E.~A.,  2001, \mn@doi
  [\apj] {10.1086/323863}, 562, 852

\bibitem[\protect\citeauthoryear{{Heyer}, {Krawczyk}, {Duval}  \&
  {Jackson}}{{Heyer} et~al.}{2009}]{Heyer+09}
{Heyer} M.,  {Krawczyk} C.,  {Duval} J.,   {Jackson} J.~M.,  2009, \mn@doi
  [\apj] {10.1088/0004-637X/699/2/1092}, 699, 1092

\bibitem[\protect\citeauthoryear{{Hughes} et~al.,}{{Hughes}
  et~al.}{2013}]{Hughes+13}
{Hughes} A.,  et~al., 2013, \mn@doi [\apj] {10.1088/0004-637X/779/1/46}, 779,
  46

\bibitem[\protect\citeauthoryear{{Johnstone}, {Wilson}, {Moriarty-Schieven},
  {Giannakopoulou-Creighton}  \& {Gregersen}}{{Johnstone}
  et~al.}{2000}]{Johnstone+00}
{Johnstone} D.,  {Wilson} C.~D.,  {Moriarty-Schieven} G.,
  {Giannakopoulou-Creighton} J.,   {Gregersen} E.,  2000, \mn@doi [\apjs]
  {10.1086/317379}, 131, 505

\bibitem[\protect\citeauthoryear{{Kainulainen}, {Beuther}, {Henning}  \&
  {Plume}}{{Kainulainen} et~al.}{2009}]{Kainulainen+09}
{Kainulainen} J.,  {Beuther} H.,  {Henning} T.,   {Plume} R.,  2009, \mn@doi
  [\aap] {10.1051/0004-6361/200913605}, 508, L35

\bibitem[\protect\citeauthoryear{{Kainulainen}, {Beuther}, {Banerjee},
  {Federrath}  \& {Henning}}{{Kainulainen} et~al.}{2011}]{Kainulainen+11}
{Kainulainen} J.,  {Beuther} H.,  {Banerjee} R.,  {Federrath} C.,   {Henning}
  T.,  2011, \mn@doi [\aap] {10.1051/0004-6361/201016383}, 530, A64

\bibitem[\protect\citeauthoryear{{Kauffmann}, {Pillai}, {Shetty}, {Myers}  \&
  {Goodman}}{{Kauffmann} et~al.}{2010a}]{Kauffmann+10a}
{Kauffmann} J.,  {Pillai} T.,  {Shetty} R.,  {Myers} P.~C.,   {Goodman} A.~A.,
  2010a, \mn@doi [\apj] {10.1088/0004-637X/712/2/1137}, 712, 1137

\bibitem[\protect\citeauthoryear{{Kauffmann}, {Pillai}, {Shetty}, {Myers}  \&
  {Goodman}}{{Kauffmann} et~al.}{2010b}]{Kauffmann+10b}
{Kauffmann} J.,  {Pillai} T.,  {Shetty} R.,  {Myers} P.~C.,   {Goodman} A.~A.,
  2010b, \mn@doi [\apj] {10.1088/0004-637X/716/1/433}, \href
  {https://ui.adsabs.harvard.edu/abs/2010ApJ...716..433K} {716, 433}

\bibitem[\protect\citeauthoryear{{Kauffmann}, {Pillai}  \&
  {Goldsmith}}{{Kauffmann} et~al.}{2013}]{Kauffmann+13}
{Kauffmann} J.,  {Pillai} T.,   {Goldsmith} P.~F.,  2013, \mn@doi [\apj]
  {10.1088/0004-637X/779/2/185}, 779, 185

\bibitem[\protect\citeauthoryear{{Kirk} et~al.,}{{Kirk} et~al.}{2017}]{Kirk+17}
{Kirk} H.,  et~al., 2017, \mn@doi [\apj] {10.3847/1538-4357/aa8631}, 846, 144

\bibitem[\protect\citeauthoryear{{Koda}, {Scoville}  \& {Heyer}}{{Koda}
  et~al.}{2016}]{Koda+16}
{Koda} J.,  {Scoville} N.,   {Heyer} M.,  2016, \apj, 823, 76

\bibitem[\protect\citeauthoryear{{K{\"o}nyves} et~al.,}{{K{\"o}nyves}
  et~al.}{2015}]{Konyves+15}
{K{\"o}nyves} V.,  et~al., 2015, \mn@doi [\aap] {10.1051/0004-6361/201525861},
  584, A91

\bibitem[\protect\citeauthoryear{{Kritsuk}, {Norman}  \& {Wagner}}{{Kritsuk}
  et~al.}{2011}]{Kritsuk+11}
{Kritsuk} A.~G.,  {Norman} M.~L.,   {Wagner} R.,  2011, \mn@doi [\apj]
  {10.1088/2041-8205/727/1/L20}, 727, L20

\bibitem[\protect\citeauthoryear{{Lada}, {Muench}, {Rathborne}, {Alves}  \&
  {Lombardi}}{{Lada} et~al.}{2008}]{Lada+08}
{Lada} C.~J.,  {Muench} A.~A.,  {Rathborne} J.,  {Alves} J.~F.,   {Lombardi}
  M.,  2008, \mn@doi [\apj] {10.1086/523837}, 672, 410

\bibitem[\protect\citeauthoryear{Larson}{Larson}{1981}]{Larson81}
Larson R.~B.,  1981, Mon. Not. R. Astron. Soc., 194, 809

\bibitem[\protect\citeauthoryear{Lombardi, Alves  \& Lada}{Lombardi
  et~al.}{2010}]{Lombardi+10}
Lombardi M.,  Alves J.~F.,   Lada C.~J.,  2010, Astron. Astrophys., 519, L7

\bibitem[\protect\citeauthoryear{{Loren}}{{Loren}}{1989}]{Loren89}
{Loren} R.~B.,  1989, \apj, 338, 902

\bibitem[\protect\citeauthoryear{Miville-Desch{\^{e}}nes, Murray  \&
  Lee}{Miville-Desch{\^{e}}nes et~al.}{2017}]{Miville-Deschenes+17}
Miville-Desch{\^{e}}nes M.-A.,  Murray N.,   Lee E.~J.,  2017, Astrophys. J.,
  834, 1

\bibitem[\protect\citeauthoryear{{Onishi} et~al.,}{{Onishi}
  et~al.}{1999}]{Onishi+99}
{Onishi} T.,  et~al., 1999, \mn@doi [\pasj] {10.1093/pasj/51.6.871}, 51, 871

\bibitem[\protect\citeauthoryear{Roman-Duval, Jackson, Heyer, Rathborne  \&
  Simon}{Roman-Duval et~al.}{2010}]{Roman-Duval+10}
Roman-Duval J.,  Jackson J.~M.,  Heyer M.,  Rathborne J.,   Simon R.,  2010,
  Astrophys. J., 723, 492

\bibitem[\protect\citeauthoryear{{Rom{\'a}n-Z{\'u}{\~n}iga}, {Alves}, {Lada}
  \& {Lombardi}}{{Rom{\'a}n-Z{\'u}{\~n}iga} et~al.}{2010}]{RomanZuniga+10}
{Rom{\'a}n-Z{\'u}{\~n}iga} C.~G.,  {Alves} J.~F.,  {Lada} C.~J.,   {Lombardi}
  M.,  2010, \mn@doi [\apj] {10.1088/0004-637X/725/2/2232}, 725, 2232

\bibitem[\protect\citeauthoryear{{Sanhueza}, {Contreras}, {Wu}  \&
  {Jackson}}{{Sanhueza} et~al.}{2019}]{Sanhueza+19}
{Sanhueza} P.,  {Contreras} Y.,  {Wu} B.,   {Jackson} J.~M.,  2019, \apj

\bibitem[\protect\citeauthoryear{{Scalo}}{{Scalo}}{1990}]{Scalo90}
{Scalo} J.,  1990, in {Capuzzo-Dolcetta} R.,  {Chiosi} C.,   {di Fazio} A.,
  eds,  Astrophysics and Space Science Library Vol. 162, Physical Processes in
  Fragmentation and Star Formation. pp 151--176,
  \mn@doi{10.1007/978-94-009-0605-1_12}

\bibitem[\protect\citeauthoryear{{Shetty}, {Collins}, {Kauffmann}, {Goodman},
  {Rosolowsky}  \& {Norman}}{{Shetty} et~al.}{2010}]{Shetty+10}
{Shetty} R.,  {Collins} D.~C.,  {Kauffmann} J.,  {Goodman} A.~A.,  {Rosolowsky}
  E.~W.,   {Norman} M.~L.,  2010, \mn@doi [\apj]
  {10.1088/0004-637X/712/2/1049}, 712, 1049

\bibitem[\protect\citeauthoryear{{Solomon}, {Rivolo}, {Barrett}  \&
  {Yahil}}{{Solomon} et~al.}{1987}]{Solomon+87}
{Solomon} P.~M.,  {Rivolo} A.~R.,  {Barrett} J.,   {Yahil} A.,  1987, \mn@doi
  [\apj] {10.1086/165493}, 319, 730

\bibitem[\protect\citeauthoryear{{Sun} et~al.,}{{Sun} et~al.}{2018}]{Sun+18}
{Sun} J.,  et~al., 2018, \mn@doi [\apj] {10.3847/1538-4357/aac326}, 860, 172

\bibitem[\protect\citeauthoryear{{Tassis}, {Christie}, {Urban}, {Pineda},
  {Mouschovias}, {Yorke}  \& {Martel}}{{Tassis} et~al.}{2010}]{Tassis+10}
{Tassis} K.,  {Christie} D.~A.,  {Urban} A.,  {Pineda} J.~L.,  {Mouschovias}
  T.~C.,  {Yorke} H.~W.,   {Martel} H.,  2010, \mn@doi [\mnras]
  {10.1111/j.1365-2966.2010.17181.x}, 408, 1089

\bibitem[\protect\citeauthoryear{{Utomo} et~al.,}{{Utomo}
  et~al.}{2018}]{Utomo+18}
{Utomo} D.,  et~al., 2018, \mn@doi [\apjl] {10.3847/2041-8213/aacf8f}, 861, L18

\bibitem[\protect\citeauthoryear{{V{\'a}zquez-Semadeni} \&
  {García}}{{V{\'a}zquez-Semadeni} \& {García}}{2001}]{VSGarcia01}
{V{\'a}zquez-Semadeni} E.,  {García} N.,  2001, \mn@doi [\apj]
  {10.1086/321688}, 557, 727

\bibitem[\protect\citeauthoryear{{V{\'a}zquez-Semadeni}, {G{\'o}mez},
  {Jappsen}, {Ballesteros-Paredes}, {Gonz{\'a}lez}  \&
  {Klessen}}{{V{\'a}zquez-Semadeni} et~al.}{2007}]{VS+07}
{V{\'a}zquez-Semadeni} E.,  {G{\'o}mez} G.~C.,  {Jappsen} A.~K.,
  {Ballesteros-Paredes} J.,  {Gonz{\'a}lez} R.~F.,   {Klessen} R.~S.,  2007,
  \mn@doi [\apj] {10.1086/510771}, 657, 870

\bibitem[\protect\citeauthoryear{{Veltchev}, {Ossenkopf-Okada}, {Stanchev},
  {Schneider}, {Donkov}  \& {Klessen}}{{Veltchev} et~al.}{2018}]{Veltchev+18}
{Veltchev} T.~V.,  {Ossenkopf-Okada} V.,  {Stanchev} O.,  {Schneider} N.,
  {Donkov} S.,   {Klessen} R.~S.,  2018, \mn@doi [\mnras]
  {10.1093/mnras/stx3267}, 475, 2215

\bibitem[\protect\citeauthoryear{{Williams}, {de Geus}  \& {Blitz}}{{Williams}
  et~al.}{1994}]{Williams+94}
{Williams} J.~P.,  {de Geus} E.~J.,   {Blitz} L.,  1994, \mn@doi [\apj]
  {10.1086/174279}, 428, 693

\bibitem[\protect\citeauthoryear{{Zamora-Avil{\'e}s}, {Ballesteros-Paredes}  \&
  {Hartmann}}{{Zamora-Avil{\'e}s} et~al.}{2017}]{ZA+17}
{Zamora-Avil{\'e}s} M.,  {Ballesteros-Paredes} J.,   {Hartmann} L.~W.,  2017,
  \mn@doi [\mnras] {10.1093/mnras/stx1995}, 472, 647

\makeatother
\end{thebibliography}

%




\bsp	
\label{lastpage}
\end{document}